\documentclass[paper]{JHEP}
\usepackage{epsfig}

\title{Paramagnetic effect of light quark loops
on Chiral Symmetry Breaking\thanks{
Work partly supported by the EEC, TMR-CT98-0169, EURODAPHNE network.}
}
\author{S.~Descotes, L.~Girlanda
and J.~Stern\\
I.P.N., Groupe de Physique Th\'eorique\\
Universit\'e de Paris-Sud, F-91406 Orsay Cedex, France\\
E-mail: \email{descotes@ipno.in2p3.fr}, \email{girlanda@ipno.in2p3.fr},
\email{stern@ipno.in2p3.fr}}
\preprint{IPNO/DR 99-27}
\abstract{We argue that light quark loops produce a paramagnetic suppression of
infrared-sensitive order parameters such as $\langle \bar q q \rangle$, as
the number $N_f$ of light fermions increases.
The possibly strong dependence of $\langle \bar q q \rangle$ on $N_f$ is
related to the observed Zweig rule violation in the scalar channel.
Presuming the existence of a chiral phase transition for not too large $N_f$,
we discuss the phenomenological possibilities of separately determining 
the two-flavour and three-flavour condensates and the quark mass
ratio $r=2 m_s/ (m_u + m_d)$. 
The issue is closely related to the interpretation of new
forthcoming  precise  $\pi\pi$ data at low energy.}


\begin{document}

\section{Introduction}

The purpose of this paper is to reconsider our understanding of
the mechanism of spontaneous chiral symmetry breaking (S$\chi$SB)
in QCD as reflected by the (possibly small) size of $\bar q q$ condensates
and to interpret its relation to the Zweig rule (ZR) violation in the scalar
channel and to the quark mass ratio  $r =2 m_s/(m_u + m_d)$. New precise
experimental results which are expected to come soon \cite{daphne,mainz,exper}
will merely concern
low-energy $\pi \pi$ scattering which is governed by the chiral dynamics
of $u$- and $d$-quarks with the $s$-quark playing
at most the role of a ``sea-side spectator''. From this point of view the
existence of a relationship between low-energy $\pi \pi$ observables and the
quark mass ratio $r =m_s/m$ ($m=m_u=m_d$) is by no means obvious. A related
question concerns determinations of $\mathrm{SU}(2)\times \mathrm{SU}(2)$ low-energy constants
from $\mathrm{SU}(3)\times \mathrm{SU}(3)$ observables: for example, the low energy constants
$l_4$ and $l_3$, which are essential to assess the prediction of standard chiral
perturbation theory (S$\chi$PT) \cite{Gasser-Leutwyler} for 
$\pi \pi$ s-wave scattering lengths \cite{pipistd}, are usually inferred using the 
experimental values of $F_K/F_{\pi}$ and of the $K$- and $\eta$-meson
masses respectively.

The theoretical counterpart of these and similar phenomenological questions
concerns the dependence of various order parameters of S$\chi$SB on the number
$N_f$ of massless quark flavours. This $N_f$-dependence is an effect of
light quark loops and it is generally expected to be rather weak: it is
suppressed in the large-$N_c$ limit and it violates the Zweig rule which
are both considered as a good approximation to the real QCD (with a possible
exception of the anomalous $0^{-+}$ channel). On the other hand, more recent
investigations suggest that $N_f$-dependent light-quark loop effects
could sometimes be rather important. First, Nature does not seem to
respect the large-$N_c$ predictions in the scalar channel     
which is not dominated by ideally mixed $\bar qq$ nonets, as required by the
Zweig rule \cite{Pennington}. Next, some recent lattice simulations with
dynamical fermions \cite{lattflav1,lattflav2}
observe a rather strong $N_f$-dependence of S$\chi$SB signals.         
Finally, a new method of estimating the variation of the chiral condensate 
between $N_f=2$ and $N_f=3$ from the
data has recently been proposed \cite{Moussallam}. 
Using as input experimental informations
on ZR violation in the scalar channel at low and medium energies,
a large variation of $\langle\bar q q\rangle$ has been found.

For sufficiently large $N_f/N_c$,
the existence of chiral phase transitions is generally expected on the
basis of the behaviour of the perturbative QCD
$\beta$-function \cite{Banks-Zaks}. Various approaches have been proposed to
study the transitions arising when $N_f$ increases. Investigations about the
QCD conformal window, where the theory is asymptotically free in the 
ultraviolet, but is governed by a non-trivial fixed point in the infrared, 
suggest a restoration of chiral symmetry for $N_f\sim 10$ at $N_c=3$
\cite{conform}. The analysis of gap equations in the same conformal window
puts a slightly different bound, with a critical $N_f$ above 12 \cite{gapeq}.
On the other hand, the instanton liquid model indicates that instantons do not 
significantly contribute to Chiral Symmetry Breaking for $N_f > 6$
\cite{instanton}.

We would like to stress that particular properties of
vector-like gauge theories such as QCD  allow a new type of 
non-perturbative interpretation  of S$\chi$SB and of its $N_f$-dependence,
suggesting a natural link between a suppression of $\langle \bar q q\rangle$ 
and an enhancement of ZR violation in the scalar channel for not too
large $N_f/N_c$. No reference is made to concepts and methods such as gap 
equation, mean field approximation, or the critical effective
coupling, which in confining non-Abelian gauge theories lack a truly 
non-perturbative gauge-invariant definition. 
Instead, QCD may be formulated in Euclidean space-time and the mechanism of
S$\chi$SB can be related to the dynamics of the lowest modes of the 
hermitean Dirac operator \cite{Banks-Casher,Vafa-Witten,Leutwyler-Smilga}
\begin{equation} \label{dirac}
H[G] = \gamma_{\mu} (\partial_{\mu} + iG_{\mu}) = H^{\dagger}[G],
\end{equation}
averaged over gauge field configurations $G_{\mu}(x)$. In particular,
the role of increasing $N_f$ naturally appears as a consequence of the
paramagnetic effect of light quark loops: it tends to suppress the order
parameters that are dominated by the small eigenvalues of $H$. Whether this
suppression is strong enough to imply a phase transition for not too large
$N_f$ is a dynamical question which at present we are not able to answer
analytically. At least one can arrive at a theoretically coherent framework
suggesting a semi-quantitative understanding of various aspects of S$\chi$SB
and of their possible interconnection.

\section{$N_f$-dependence of infrared-dominated order parameters}

The theory will be considered in a four-dimensional
Euclidean box $L \times L \times L \times L$ with (anti)periodic
boundary conditions modulo a gauge transformation (i.e. on a torus).
We are interested in gauge invariant correlation functions which break the
chiral symmetry of the vacuum, i.e. in vacuum expectations  of operators
which do not contain the singlet representation of the chiral symmetry group.
All external momenta are set to 0 and all quarks are first taken to be
massive. At the end we consider the limit in which the first $N_f$ lightest
quarks become massless keeping the remaining quark masses fixed. 
One randomly chooses a gauge field configuration
$G_{\mu}(x)$ and performs the Grassmann integral over quark fields.
The result can be formally expressed in terms of eigenvalues and
eigenfunctions of the hermitean Dirac operator (\ref{dirac}):
\begin{equation} \label{eigen}
H[G] \phi_n = \lambda_n[G] \phi_n, \qquad\int dx\ \phi_n^{\dagger}(x) \phi_k(x)
=\delta_{nk}.
\end{equation}
Since $H[G]$ anticommutes with $\gamma_5 =\gamma_5^{\dagger}$, the spectrum
is symmetric around $0$. In addition, for gauge field configurations with
non-vanishing winding number $\nu \in Z$, the spectrum contains $|\nu|$
topological zero modes. Positive eigenvalues will be ranged
in the ascending order and numerated by a positive integer, defining
$\lambda_{-n}=-\lambda_{n}$. Notice that the eigenvalues $\lambda_{n}$  and
the wave functions
$\phi_{n}(x)$ are entirely given by the gauge field configuration
$G_{\mu}(x)$;
in particular, they are independent of the quark flavour and of the quark
masses\footnote{The wave functions $\phi_{n}$ live in the spin $\times$ colour      
space and they transform as the fundamental representation of
$\mathrm{SU}(N_c)$ and as a 4-dimensional 
$\mathrm{O}(4)$ spinor respectively.}.
The dependence on quark flavour and masses arises from the propagator
\begin{equation} \label{prop}
S_{j}(x,y|G) =\sum_{n}\frac{\phi_{n}(x)\phi_{n}^{\dagger}(y)}
{m_j - i\lambda_{n}[G]} \end{equation}
and from the fermionic determinant $\det(M-iH)=\prod_{j}\Delta(m_j|G)$, where
the single flavour determinant can be expressed as
\begin{equation}\label{det}
\Delta(m|G) =m^{|\nu|} \prod _{n > 0}\frac{m^2 + \lambda_{n}^2[G]}{m^2 +
\omega_{n}^2}.\end{equation}
The $G$-independent numbers $\omega_{n}$, which essentially coincide with the
free eigenvalues, provide a convenient overall
normalization of (\ref{det}) and will be specified shortly. An integral over
fermion fields of any product of bilinear quark currents can be expressed
in terms of the propagator (\ref{prop}) and of the determinant (\ref{det}). 
The next step then consists in taking an average over all gluon 
configurations. Before
we comment on this last point, let us concentrate on the simplest example  
of an order parameter of S$\chi$SB.

The chiral condensate $\langle\bar u u\rangle$, where $u$ stands for 
the lightest quark,
will be considered in the $\mathrm{SU}(N_f)\times \mathrm{SU}(N_f)$ chiral limit
\begin{equation} \label{lim}
m_1 = m_2 = \ldots = m_{N_f} = m \rightarrow 0, \qquad  m_j \neq 0,\ j>N_f
\end{equation}
and it will be denoted as $-\Sigma(N_f)$. One has
\begin{equation} \label{cond}
\Sigma (N_f) = \lim \frac{1}{L^4} \ll \int\ dx
{\mathrm{Tr}}\ S(x,x|G)\gg _{N_f} = \lim \frac{1}{L^4} \ll
  \sum_{n} \frac{m}{m^2 + \lambda_{n}^2}\gg _{N_f}.
\end{equation}
Hereafter, $\lim$ denotes the $\mathrm{SU}(N_f)\times \mathrm{SU}(N_f)$ limit (\ref{lim})
preceeded by the large volume
limit. The symbol $\ll \gg _{N_f}$ represents the normalized
($\ll\!1\!\gg_{N_f}=1$) average over gauge-field configurations weighted by
the fermionic determinant,
\begin{equation} \label{average}
\ll \Gamma \gg _{N_f} = Z^{-1} \int d\mu[G]\ \Gamma\ \Delta^{N_f}(m|G)   
\prod _{j>N_f}\Delta (m_j|G)  \exp\{-S[G]\},
\end{equation}
where $S[G]$ stands for the Yang-Mills action.
Since every gauge-field configuration $G_{\mu}(x)$ can be globally
characterized by the corresponding set of Dirac eigenvalues ${\lambda}$ and
eigenvectors ${\phi}$, the functional integral (\ref{average}) may be viewed
as an average over all possible Dirac spectra. The probability distribution
of Dirac eigenvalues should be, in principle, calculable from the theory
itself. In practice, it requires a non-perturbative regularization and
renormalization of the gluonic average (\ref{average}) which (as in 
perturbation theory) may depend on the observable $\Gamma$. Whilst
this problem is hard to solve in general, in the particular case of chiral
order parameters such as (\ref{cond}), the formal structure of
Eq.~(\ref{average}) suggests some possibly interesting properties of
S$\chi$SB even before an analytic solution becomes available.

The fact that $\langle\bar u u\rangle$ is an order parameter of S$\chi$SB is
reflected by the vanishing of the $m \to 0$ limit of Eq.~(\ref{cond})
taken at finite volume. In order to get a non-trivial result, the large
volume limit has to be taken first and the spectrum of the Dirac operator
has to become sufficiently dense around the origin. Actually, the average
distribution of the smallest Dirac eigenvalues is all what matters: if in 
Eq.~(\ref{cond}) one cuts the infrared end of the spectrum and sticks to
$|\lambda_n|>\Lambda$, the result would be zero, no matter how small
$\Lambda$ is. For the same reason, the ultraviolet divergences of the sum 
$\sum_{n}$
in Eq.~(\ref{cond}) become irrelevant in the chiral limit. It turns out
that only eigenvalues which in the average accumulate like $1/L^4$ contribute
to the chiral condensate $\langle\bar q q\rangle$
\cite{Banks-Casher,Vafa-Witten,Leutwyler-Smilga}.
A similar discussion applies to other
order parameters of S$\chi$SB, in particular to the coupling $F_{\pi}$ of
Goldstone bosons to the axial current, whose square can be expressed as
the two-point correlator of left-handed and right-handed currents
$\langle L_{\mu}R_{\nu}\rangle$ at zero momentum transfer. 
Denoting by $F^2 (N_f)$ the $\mathrm{SU}(N_f)\times \mathrm{SU}(N_f)$ 
chiral limit (\ref{lim}) of $F^2 _{\pi}$, one has
\cite{Stern}
\begin{equation} \label{conduct}
F^2 (N_f) = \lim \frac{1}{L^4} \ll \sum _{k,n} \frac {m}{m^2 + \lambda ^2 _{k}}
\frac {m}{m^2 + \lambda ^2 _{n}} J_{kn} \gg _{N_f},
\end{equation}
where $\lim$ has the same meaning as in Eq.~(\ref{cond}) and
\begin{equation} \label{mob}
J_{kn} = \frac{1}{4} \sum_{\mu} \left|\int dx\ \phi^{\dagger} _{k} (x) 
  \gamma _{\mu} \phi _{n}(x)\right|^2.
\end{equation}
Due to the Goldstone theorem, $F^{2}(N_f) \ne 0$ is both sufficient and
necessary for the chiral symmetry $\mathrm{SU}(N_f) \times \mathrm{SU}(N_f)$ 
to be spontaneously broken. Again, $F^{2}(N_f)$  merely receives 
contributions
from the lowest Dirac eigenvalues but it is less infrared sensitive
than $\langle\bar u u\rangle$: the eigenvalues behaving in the average as
$1/L^2$ could now be sufficient to produce a non-zero value of
(\ref{conduct}), since there are two factors 
$m/(m^2 + \lambda^2)$ for a single inverse power of volume \cite{Stern}.

Eq.~(\ref{average}) suggests that for $m \to 0$ the effect of the fermionic
determinant and the $N_f$-dependence will be stronger for observables
$\Gamma$ which are dominated by the lowest Dirac eigenvalues. This observation
follows from the rigorously proven inequality~\cite{Vafa-Witten}
\begin{equation} \label{VW}
|\lambda _{n}[G]| < C \frac{n^{1/d}}{L} \equiv \omega _{n},
\end{equation}
where $d$ is the space-time dimension ($d=4$ in our case) and $C$ is a constant
independent of the gauge field configuration $G_{\mu}(x)$, of the integer $n$
and of the volume $V=L^d$. (In general, $C$ depends on the shape of the space
time manifold, once the volume has been fixed.)
The existence of the uniform upper bound (\ref{VW}) reflects
the paramagnetic response of the Euclidean Dirac spectrum to an external
gauge field \cite{paramagnet}.
It allows to split the single flavour determinant (\ref{det})
into infrared and ultraviolet parts \cite{Duncan-Eichten}:
one chooses a cutoff $\Lambda$ and
defines an integer $K$ such that $\omega _{K} = \Lambda$. The determinant is
then written
\begin{equation} \label{split}
\Delta(m|G) = m^{|\nu|} \Delta_\mathrm{IR}(m|G) \Delta_\mathrm{UV}(m|G),
\end{equation}
where $\Delta_\mathrm{IR}$ involves the first $K$ non-zero eigenvalues and is
bounded by 1 as a  consequence of the inequality (\ref{VW}),
\begin{equation} \label{IR}
\Delta_\mathrm{IR}(m|G) = \prod \limits_{k=1}^{K} \frac{m^2 + \lambda_{k}^{2}[G]}
{m^2 +\omega _{k}^2} < 1. \end{equation}
One may expect that in the case of chiral order parameters such as
$\langle\bar u u\rangle$
(\ref{cond}) or $F^2$ (\ref{conduct}) which are entirely dominated by the
infrared extremity of the Dirac spectrum, the $\Delta_\mathrm{IR}$ part of the
determinant will describe the bulk of the effect of light quark loops in
Eq.~(\ref{average}). This effect should be paramagnetic,
\begin{equation} \label{para}
\Sigma (N_f + 1) < \Sigma (N_f),\qquad F^2 (N_f + 1) < F^2 (N_f),
\end{equation}
indicating the suppression of chiral order parameters with increasing $N_f$.
How strong is this suppression depends on how sensitive is the observable
$\Gamma$ in Eq.~(\ref{average}) to the smallest Dirac eigenvalues for large
volumes and small quark masses. For this reason one may expect a stronger
suppression in the case of $\langle \bar q q\rangle$ than for $F_{\pi}$.
On the other hand, for observables that are not dominated by the
infrared extremity of the Dirac spectrum, the sensitivity to the
determinant and to $N_f$ can remain marginal, as expected in the large-$N_c$
limit.

\section{Why the scalar channel does not obey large-$N_c$ predictions}

We now turn to the connection between flavour dependence of order
parameters of S$\chi$SB and the observed violation of the Zweig rule in
the scalar channel. As before, we consider first $N_f$ light flavors of
common mass $m \to 0$ and denote by $s$ the ($N_{f}+1$)-th quark, whose mass
$m_s$ is non-zero, but still considered as light compared to the scale of the
theory. For $N_{f}=2$, this corresponds to the situation in real QCD.
$\Sigma(N_f)$ is a function of $m_s$, and its derivative reads
\begin{eqnarray} \label{der}
\frac{\partial}{\partial m_s} \Sigma(N_f) &=& 
\lim_{m\to 0} \int\ dx \langle \bar u u(x) \bar s s(0)\rangle^{c} 
   \ \equiv\ \Pi_{Z}(m_s)     \nonumber \\
&=&\lim \frac{1}{L^4} \ll \left(\sum_{k} \frac{m}{m^2 + \lambda_{k}^2}\right)
  \left(\sum_{n} \frac{m_s}{m_{s}^2 + \lambda_{n}^2}\right)\gg _{N_f}^c ,
\end{eqnarray}
where the notations are as before and the superscript $c$ stands for the
connected part. Since $\Sigma(N_f) \to \Sigma(N_{f}+1)$
for $m_s \to 0$, one can write
\begin{equation} \label{diff}
\Sigma(N_f) = \Sigma(N_{f}+1) + \int_{0}^{m_s}\!\!d\mu\ \Pi_{Z}(\mu)=
\Sigma(N_{f}+1) + m_s Z_\mathrm{eff}^S(m_s) + O(m_{s}^2 \log m_s).
\end{equation}
In general, and for $N_f \sim 2-3$, the difference
$\Delta(N_f) = \Sigma(N_f) -\Sigma(N_{f}+1)$ is expected to be negligible
compared to $\Sigma(N_f)$ for two independent reasons: first, it is chirally
suppressed due to the smallness of $m_s$ relative to the condensate
$\Sigma(N_{f}+1)$, at least provided the latter is of the standard ``normal''
size. Second, the connected correlator (\ref{der}) of scalar quark densities
of different flavours is suppressed in the large-$N_c$ limit relative to
$\Sigma(N_f)$. An important $N_f$-dependence of the condensate would imply
that \emph{both} these arguments fail. We argue that this should naturally
be expected close to the critical point $n_{\mathrm{crit}}(N_c)$ 
at which $\Sigma(N_f)$ vanishes. Suppose that for a given value of $N_c$
(e.g. $N_c=3$), we have $N_{f} + 1 <n_{\mathrm{crit}}(N_c)$, so that
$\Sigma(N_{f}+1)$ is still non-zero but
already small. Then (for the actual value of $m_s$) the condensate term
need not dominate the expansion (\ref{diff}) in powers of $m_s$, not because
the chiral expansion breaks down but due to the suppression of
$\Sigma(N_{f}+1)$. The generalized chiral perturbation theory (G$\chi$PT) is
precisely designed to cope with such a situation
\cite{gchipt,pionpion1,pionpion2}.
The suppression of the
condensate means that near the critical point, the average
density of states \cite{Banks-Casher}
\begin{equation} \label{density}
\frac{1}{L^4} \ll \rho (\lambda)\gg _{N_f}, \qquad \rho (\lambda) =
\sum_{n}\delta(\lambda -\lambda_{n}[G]),
\end{equation}
drops for $\lambda \sim m$. Indeed, Eq.~(\ref{cond}) can be rewritten as
\begin{equation} \label{BC}
\Sigma(N_f) = 2 \lim \int_{0}^{\infty} \frac{du}{1+u^2} \frac{1}{L^4}
\ll \rho (mu)\gg_{N_f}\!.
\end{equation}
It is then natural to expect that  the proximity of
a phase transition will further manifest itself by an increase of
fluctuations of the density of states and/or by an enhancement of the
density-density correlation
$L^{-4}\ll\!\rho(\lambda)\rho(\lambda^{\prime})\!\gg^{c}_{N_f}$ for
$\lambda \sim \lambda^{\prime} \sim m$.  This quantity determines the
variation of the condensate, see Eqs.~(\ref{der}) and (\ref{diff}):
\begin{equation} 
\frac{\partial}{\partial m_s} \Sigma(N_f) =
    4 \lim \int_{0}^{\infty} \frac{du}{1+u^2}
\frac{dv}{1+v^2} \frac{1}{L^4} \ll \rho(mu) \rho(m_{s}v)\gg^{c}_{N_f}\!.
\end{equation}
Far away from the critical point, the correlation of small Dirac eigenvalues
should not be important, as predicted by the large-$N_c$ limit.
Since this limit leads to a suppression of quark loops,
any infrared-dominated order parameter becomes independent of $N_f$.
Therefore, the large-$N_c$ limit prevents $\Sigma$ from vanishing, since
the critical number of flavours $n_\mathrm{crit}(N_c)$ moves away to
infinity as  $N_c\to\infty$. This asymptotic behaviour is
also supported by perturbative calculations,
in which $N_f$ and $N_c$ usually arise through their ratio.
Large-$N_c$ expansion is therefore expected to converge slowly for $N_f$
fixed just below the critical point $n_{\mathrm{crit}}(N_c)$, and to yield
irreparably false results for $N_f$ above it.
This argument merely concerns the vacuum channel $0^{++}$. The variation of
any chiral order parameter between $N_f$ and $N_f+1$ is given by a
correlation function which violates the Zweig rule precisely in that
channel, cf Eq.~(\ref{diff}): a strong variation would imply the existence
of $J^{PC}=0^{++}$, SU$_V(N_f)$-singlet states strongly coupled both to the
first $N_f$ light quarks and to the \emph{scalar} density $\bar s s $
of the $(N_f+1)$-th quark. The proximity of a phase transition could then
explain in a natural way why the spectrum of $0^{++}$ states is not
dominated by ideally mixed scalar mesons, presenting significant
discrepancies with large-$N_c$ predictions\footnote{
Connections with scenarios invoking trace anomaly and light
dilatons~\cite{dilaton} remain
to be seen. We thank H.~Leutwyler for bringing our attention to this
question.}.

In the following we concentrate on the actual case $N_f = 2-3$,
having in mind the possibility that (for $N_c=3$)
the real world might already be close to the critical point. It has recently
been pointed out \cite{Moussallam} that the correlation function
(\ref{der}) satisfies a
well convergent sum rule which allows a phenomenological estimate of
the $N_f$-dependence of the condensate between $N_f=2$ and $N_f=3$,
\begin{equation} \label{sumrule}
\Pi_{Z}(m_s) = \frac{1}{\pi} \int_{0}^{\infty}\frac{dt}{t}\ \sigma(t),
\end{equation}
where the spectral function $\sigma(p^2)$ (defined in  Minkowski
space-time) collects ZR violating contributions which couple both
to $\bar uu$ and $\bar ss$
\begin{equation} \label{spectral}
\sigma(p^2) = \frac{1}{2}\sum_{n} (2\pi)^{4} \delta^{(4)}(p - P_{n})\langle 0|\bar
uu|n\rangle \langle n|\bar ss|0\rangle.   
\end{equation}
Since we take $m_u = m_d=m$ and isospin symmetry cannot be spontaneously
broken \cite{restbreak}, only isoscalar states $|n\rangle$ contribute in
Eq.~(\ref{spectral}) and $\langle 0 | \bar u u | n \rangle = \langle 0 |
\bar d d | n \rangle$.
Under the plausible assumption that the dominant contribution comes from
two-particle states ($|\pi\pi\rangle, |K \bar K\rangle, \ldots $),
the spectral function
(\ref{spectral}) can be reconstructed from the corresponding multi-channel
T-matrix, which contains the experimental information on the size of Zweig
rule violation in the $0^{++}$ channel, such as the effect of the $f_0(980)$
resonance. Using further $\chi$PT to normalize the
solution of the multi-channel Muskhelishvili--Omn\`es equation, one can obtain
the corresponding form factors $\langle 0|\bar q q |n\rangle$. Within
the standard version of $\chi$PT, a strong $N_f$-dependence of the condensate
has been found \cite{Moussallam}: $\Sigma(3)/\Sigma(2) = 1-0.54\pm 0.27$. 
We shall further comment on this result shortly.
 
\section{$\chi$PT considerations}

The case of a large Zweig-rule violation leading to a substantial difference
between the $N_f=2$ and $N_f=3$ condensates has never been fully
included into the $\chi$PT analysis before. This leads us to
carefully reconsider the G$\chi$PT relation between the actual size of the
condensate(s), the quark mass ratio $r=m_s/m$, and some low
energy observables. What matters is the renormalization group invariant product
$m\Sigma$ in physical units of $F^{2}_{\pi} M^{2}_{\pi}$, i.e. 
the Gell-Mann--Oakes--Renner ratio(s) \cite{gor}
\begin{equation} \label{GOR}
X(N_f) = \frac{2m \Sigma(N_f)}{F^{2}_{\pi} M^{2}_{\pi}}.
\end{equation}
The standard chiral expansion presumes that both $X(2)$ and $X(3)$ are close to
1 which is, a priori, hardly compatible with an important flavour dependence
of the condensate. On the other hand,
in G$\chi$PT, the GOR ratio (\ref{GOR}) can be significantly below 1, 
leaving enough space for a large
$N_f$-dependence. It is worth stressing that an important ZR violation
would not affect the quantitative relation between $X(2)$ and low energy
$\pi\pi$ observables provided it is systematically based on 
$\mathrm{SU}(2)\times \mathrm{SU}(2)$ G$\chi$PT. The
reason is not a ``bad convergence'' of the expansion in the strange quark
mass. The ZR violation effects do not manifest themselves through a
specific low-energy constants in the $N_f=2$ effective Lagrangian, whereas
in the three-flavour $\mathcal{L}_\mathrm{eff}$ 
they show up as extra (ZR violating) low energy
constants $L_{6}(\mu)$ and $L_{4}(\mu)$ which are a priori unknown and
have never been determined experimentally\footnote{The reason of this
difference resides in group theory: while the ZR violating correlator
$\langle \bar uu(x) \bar dd(0)\rangle$ does not break the chiral symmetry
$\mathrm{SU}(2)\times \mathrm{SU}(2)$, the two-point function 
$\langle \bar uu(x) \bar ss(0)\rangle$ is an order parameter for both $N_f=2$
and $N_f=3$ chiral symmetry.}. Hence, $X(2)$ remains accessible to experiment
via low-energy $\pi\pi$ phases \cite{pionpion1,pionpion2}, azimuthal
asymmetries in the decay $\tau \to 3 \pi \nu_{\tau}$ \cite{tau3pi}, 
or in the reaction $\gamma+\gamma \to 3 \pi$ \cite{gammagamma}.
The question remains how this information is related to the size of $X(3)$,
to the ZR violation in the $0^{++}$ channel and to the quark
mass ratio $r=m_s/m$.

A partial answer to these questions can be obtained from a new look
at old $\mathrm{SU}(3)\times \mathrm{SU}(3)$ G$\chi$PT expansions of
kaon and pion masses (see Ref.~\cite{gchipt} and App.~A of
Ref.~\cite{pionpion2}), 
rewritten as
\begin{eqnarray}
F^2_\pi M^2_\pi &=& 2m\left[\Sigma(3)+(m_s+2m) Z_\mathrm{eff}^S(m_s)\right]
     +4m^2 A_\mathrm{eff} + F^2_\pi \delta M^2_\pi, 
   \label{expan1}\\
F^2_K M^2_K &=& (m+m_s)\left[\Sigma(3)+(m_s+2m) Z_\mathrm{eff}^S(m_s)\right]
   \nonumber\\
&&\qquad    +(m+m_s)^2 A_\mathrm{eff} + F^2_K \delta M^2_K.\label{expan2}
\end{eqnarray}
These formulae collect in a scale independent manner all contributions
linear {\em and} quadratic in quark masses. They are useful
to the extent that $\delta M^2_{P} \ll M^2_{P}$, which is certainly a much
weaker requirement than the assumption which is at the basis of S$\chi$PT.
The constants $Z_\mathrm{eff}^S$ and $A_\mathrm{eff}$ 
are related to the quark-mass independent constants of the $O(p^2)$ G$\chi$PT
Lagrangian, $Z_0^S$ and $A_0$, renormalized at a scale $\mu$:
\begin{eqnarray}
Z_\mathrm{eff}^S(m_s)&=&2F^2(3) Z_0^S(\mu)-\frac{B_0^2}{16\pi^2}
    \left\{\log\frac{\bar{M}_K^2}{\mu^2}+
        \frac{2}{9}\log\frac{\bar{M}_\eta^2}{\mu^2}\right\},
     \label{eff1}\\
A_\mathrm{eff}&=&F^2(3) A_0(\mu) \nonumber\\
&& \quad -\frac{B_0^2}{32\pi^2}
    \left\{\log\frac{\bar{M}_K^2}{\mu^2}+
        \frac{2}{3}\log\frac{\bar{M}_\eta^2}{\mu^2}
	+3\log\frac{M_K^2}{\bar M_K^2}
	+\frac{10}{9}\log\frac{M_\eta^2}{\bar M_\eta^2}
	\right\},\label{eff2}
\end{eqnarray}
where we use the notations
$B_0=\Sigma(3)/F^2(3)$, $\bar M^2_{K,\eta}=\lim_{m\to 0} M^2_{K,\eta}$.
Both expressions (\ref{eff1}) and (\ref{eff2}) are scale independent.
The connection with the S$\chi$PT $O(p^4)$ constants is
$F^2(3)Z_0^S(\mu)=16 B_0^2L_6(\mu)$ and $F^2(3)A_0(\mu)=16 B_0^2L_8(\mu)$
respectively. The constant $Z_{\mathrm{eff}}^S(m_s)$,
independent of $m_u=m_d=m$, is the same as in Eq.~(\ref{diff}), taken for
$N_f=2$. It is convenient to split the scale independent remainders
$\delta M^2_{\pi,K}$ into two scale independent parts,
$\delta M^2 = \delta_{(1)}M^2 + \delta_{(2)}M^2$:
\begin{eqnarray}
\delta_{(1)} M_\pi^2&=&\frac{4m^2 B_0^2}{32\pi^2 F_\pi^2}\nonumber\\
&&\qquad\times    \left\{3\log\frac{M_K^2}{M_\pi^2}
        +\log\frac{M_\eta^2}{M_K^2}
	+\frac{m_s}{m}\left[\log\frac{\bar M_K^2}{M_K^2}
	+\frac{2}{9}\log\frac{\bar M_\eta^2}{M_\eta^2}\right]
	\right\},\label{logs1}\\
\delta_{(1)} M_K^2&=&\frac{m(m+m_s) B_0^2}{32\pi^2 F_K^2}\nonumber\\
&&\qquad\times   \left\{3\log\frac{M_K^2}{M_\pi^2}
        +\log\frac{M_\eta^2}{M_K^2}
	+2\log\frac{\bar M_K^2}{M_K^2}
	+\frac{4}{9}\log\frac{\bar M_\eta^2}{M_\eta^2}
	\right\}.\label{logs2}
\end{eqnarray}
Substituting Eqs.~(\ref{logs1}) and (\ref{logs2}) in Eqs.~(\ref{expan1}) and
(\ref{expan2}),
one recovers the full $O(p^4)$ standard expansion \cite{Gasser-Leutwyler}.
In this case, $\delta_{(2)}M^2$ consists of
$O(p^6)$ (two-loop) and higher standard contributions \cite{Bijnens}.
In the G$\chi$PT reading, $\delta_{(2)}M^2$ are $O(m_\mathrm{quark}^3)$ as
well, but now, they consist
of the tree $\mathcal{L}_{(0,3)}$ component of
$\mathcal{L}_\mathrm{eff}$, of the remaining
scale independent part of the one-loop $O(p^4)$ contributions not included in
Eqs.~(\ref{logs1}) and (\ref{logs2}), and of higher order terms.
As a result we expect $\delta M^{2}_{\pi,K}/M^{2}_{\pi,K}$ to be at most
3-4 per cent in the whole
range $0< X(3) < 1$ (this statement will be made more quantitative in the
final result). The control of the accuracy, independently of the size of
$X(3)$, is considerably simplified expanding the product $F_P^2 M_P^2$ rather
than $M_P^2$ and $F_P^2$ separately. It avoids uncertainties related to the
low energy constant $\xi$ ($L_5$) and to its ZR violating counterpart
$\tilde \xi$ ($L_4$). Further advantages of this way of
ordering the expansion of Goldstone boson masses will appear shortly.

A simple algebra allows one to infer from Eqs.~(\ref{expan1}) and 
(\ref{expan2}) a relation between $X(3)$, the ZR violating constant
$Z_\mathrm{eff}^S$ and the quark mass ratio $r = m_s/m $:
\begin{equation} \label{x3}
X(3) + \frac{2m(m_{s}+2m)}{F^{2}_{\pi}M^{2}_{\pi}} Z_\mathrm{eff}^S = 1 - \tilde
\epsilon(r) + \delta X(3),
\end{equation}
where $\delta X(3)$ is a simple combination\footnote{
Hereafter, all manipulations with Eqs.~(\ref{expan1}) and~(\ref{expan2}) are
algebraically exact, making no use of expansions in $\delta M_{\pi,K}^2$ or
in quark masses.} of $\delta M^{2}_{\pi,K}$ and

\begin{equation} \label{epsilon}
\tilde \epsilon(r) = 2 \frac{\tilde r_{2}-r}{r^{2}-1},\qquad        
\tilde r_{2}=2\frac{F^{2}_{K}M^{2}_{K}}{F^{2}_{\pi}M^{2}_{\pi}} - 1 \sim 39.
\end{equation}
This information can now be combined with the general Eq.~(\ref{diff})
(considered here for $N_f=2$). The latter can be recovered considering the
limit $\Sigma(2) = \lim_{m \to 0}(F^{2}_{\pi}M^{2}_{\pi})/2m$ of Eq.
(\ref{expan1}) keeping $m_s$ fixed:
\begin{equation} \label{x2}
X(2) = X(3) + \frac{2mm_s}{F^{2}_{\pi}M^{2}_{\pi}} Z_\mathrm{eff}^{S} 
  +\delta X(2).
\end{equation}
Eliminating $Z_\mathrm{eff}^{S}$ from Eqs.~(\ref{x3}) and (\ref{x2}), one arrives
at a simple relation between the $N_f=2$ and $N_f=3$ GOR ratios (\ref{GOR})
and the quark mass ratio $r$
\begin{equation}\label{rel}
X(2) = [1 - \tilde \epsilon(r)] \frac{r}{r+2} + \frac{2}{r+2} X(3) + \Delta.
\end{equation}

Before we show that the remainder $\Delta$ is small and well under control,
the simple meaning of Eq.~(\ref{rel}) should be stressed. The
three-flavour GOR ratio should be in the interval $0\le X(3) \le X(2)$
because of the vacuum stability and of the paramagnetic inequality
(\ref{para}). Since, furthermore,
$r>\tilde r_{1}=2 F_{K}M_{K}/F_{\pi}M_{\pi}-1 \sim 8$,
the second term on the r.h.s.
of Eq.~(\ref{rel}) represents a small correction all over the interval of $r$:
the quark mass ratio $m_s/m$ is merely given by the two-flavour GOR ratio
$X(2)$ which is more easily accessible experimentally, whereas $X(3)$
represents the fine tuning of the above relation. It might be, for instance,
conceivable that $X(2)$ would be close to 1, the quark-mass ratio $r$ close
to its standard value (or even larger) and yet $X(3) \sim 0$, implying
a very steep decrease of $X(N_f)$ to the critical point and a huge amount
of ZR violation. This is a possibility which has never been
considered before.

We finally discuss the uncertainty in Eqs.~(\ref{x3}), (\ref{x2}) and
(\ref{rel}). One has
\begin{equation} \label{err3}
\delta X(3) = \left[\tilde \epsilon(r) +\frac{2}{r-1}\right]
   \frac{\delta M^{2}_K}{M^{2}_K}
- \frac{r+1}{r-1} \frac{\delta M^{2}_{\pi}}{M^{2}_{\pi}},
\end{equation}
whereas $\delta X(2)$ can be read off from the expansion of
$F^{2}_{\pi}M^{2}_{\pi}$ inspecting linear terms in the limit $m \to 0$
with $m_s$ fixed (including the whole $O(p^4)$ G$\chi$PT order, see 
Eq.~(5.46) in \cite{pionpion2}). As a result,
$\delta X(2) =\delta_{(2)}M^{2}_{\pi}/M^{2}_{\pi} +\ldots $, where the
dots stand for contributions for which the dimensional estimate gives
$O[m^{2}m_s/(M^{2}_{\pi}\Lambda_{H})] \sim \pm 0.4/r^2$. Hence,
this uncertainty is at the per mille level. $\Delta$ is a simple linear
combination
$\Delta = \delta X(2) + \delta X(3)\cdot r/(r+2)$ and it may be represented
as $\Delta = \Delta_{1} \pm \epsilon$. $\Delta_{1}$ represents the
contributions of chiral logs to (\ref{err3}) arising from
$\delta_{(1)}M^{2}_{\pi,K}$. The contribution of $\delta_{(2)}M^{2}_{\pi}$ to
$\Delta$ almost exactly cancels between $\delta X(3)$ and $\delta X(2)$ and
the remaining uncertainty $\epsilon$ is dominated by $\delta_{(2)}M^{2}_K$.
The dimensional estimate gives $\epsilon = 0.025$ for $r=10$, $\epsilon=0.006$
for $r=20$ and even much smaller values for higher $r$. In 
Fig.~\ref{x2vsr}, the correlation between $X(2)$ and $r$ is plotted,
including $\Delta_{1}$: Eq.~(\ref{x2}) is considered either with a maximal ZR
violation $[X(3)=0]$ or no violation at all $[X(3)=X(2)]$. We check the
fine tuning role of $X(3)$ in the significant correlation between $r$ and
$X(2)$. If the two-flavour GOR ratio is close to 1, the quark mass ratio
is not very accurately determined, but it is much more restricted for smaller
$X(2)$.

\EPSFIGURE{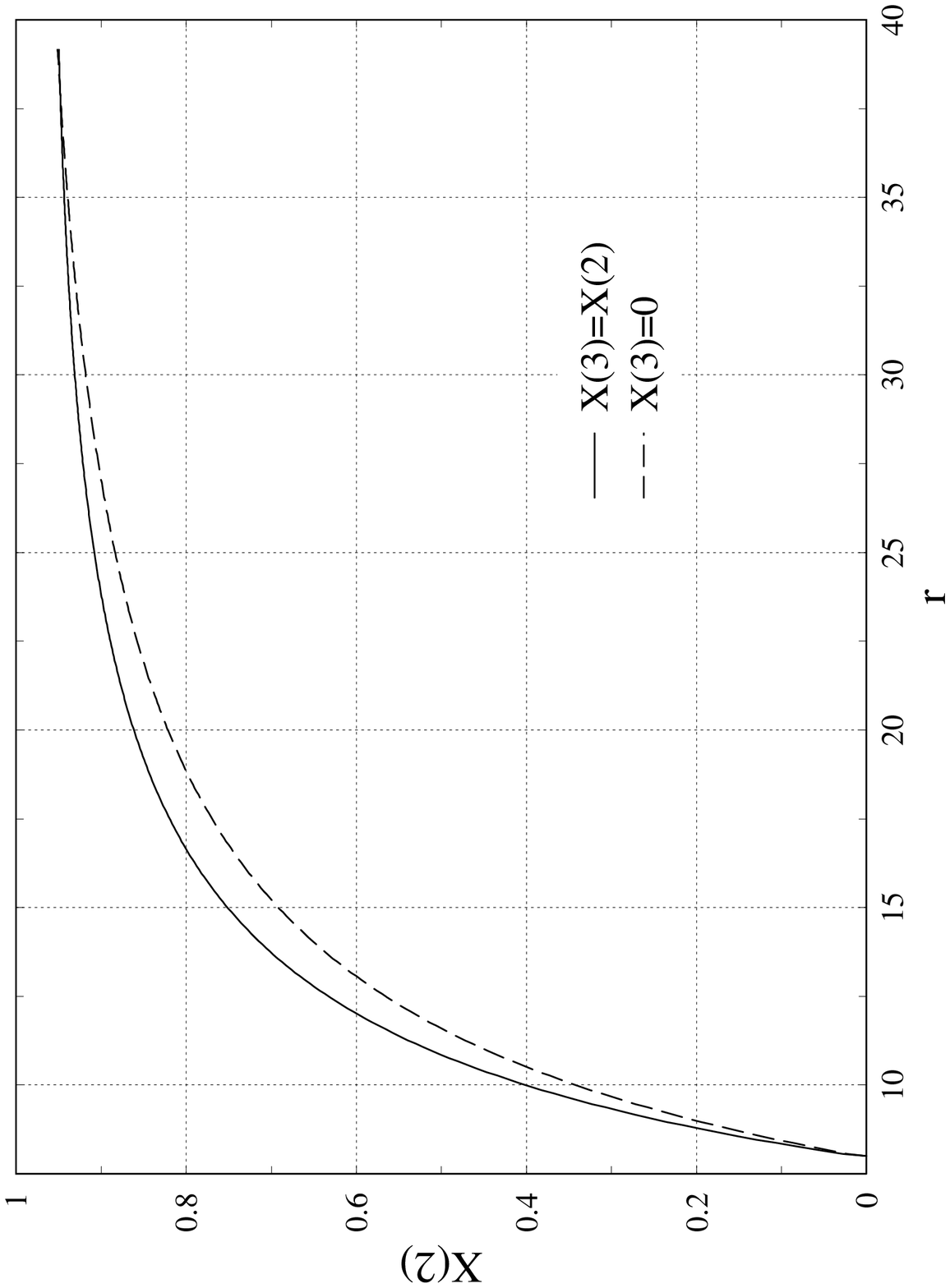,width=10cm,angle=-90}{Correlation between the
Gell-Mann--Oakes--Renner ratio for two massless flavours $X(2)$ and the
quark mass ratio $r=m_s/m$. The uncertainty due to $\epsilon$ is not taken
into account (see text). The upper curve respects strictly the Zweig rule,
which is maximally violated on the lower curve.
\label{x2vsr}}

Further insight can be obtained combining the general perturbative
expressions displayed above with Moussallam's sum rule (\ref{sumrule}).
Differentiating Eq.~(\ref{diff}) with respect to $m_s$ and using 
Eq.~(\ref{eff1}) yields the identity
\begin{equation}\label{delta}
X(2)-X(3)=\frac{2mm_s}{F^2_\pi M^2_\pi} 
   \left[\Pi_Z(m_s) + \frac{B^2_0}{16\pi^2}
     \left(\bar\lambda_K+\frac{2}{9}\bar\lambda_\eta\right)\right] + \Delta X,
     \label{diffgor}
\end{equation}
where $\bar\lambda_P=m_s\cdot\partial(\log \bar{M}_P^2)/\partial m_s$, and
\begin{equation}
\Delta X=\frac{2m}{M^2_\pi F^2_\pi}
  \left(1-m_s \frac{\partial}{\partial m_s}\right)
   \lim_{m\to 0} \frac{F^2_\pi \delta_{(2)}M^2_\pi}{2m}.
\end{equation}
The identity (\ref{diffgor}) is a variant of Eq.~(\ref{x2}), in which the
ZR violating constant $Z_\mathrm{eff}^S$ has not been eliminated
but reexpressed using the sum rule (\ref{sumrule}) taken at $m=0$. $\Delta X$
differs from $\delta X(2)$ by insertion of
$(1- m_s \partial/\partial m_s)$. They are both expected of the same
order of magnitude and of opposite sign. In particular, $\Delta X$ receives
contributions starting at the two-loop order in S$\chi$PT.
Following the same dimensional estimate as in the discussion of the
uncertainty in Eq.~(\ref{rel}), one expects
$|\Delta X | < 0.05$ for $r \sim 10$, further decreasing for larger $r$.

Eq.~(\ref{delta}) provides a general framework for the discussion of the
variation of the condensate $X(N_f)$
between $N_f=2$ and $N_f=3$ in terms of the sum rule (\ref{sumrule}),
independently of any prejudice about the size of $X(3)$. It is in particular
valid in S$\chi$PT, in which the deviation of $X(3)$ from $1$ is considered
as a small perturbation. In the latter case,
$\bar\lambda_K =\bar\lambda_{\eta} = 1$ at the leading order, and
$2m m_sB_{0}^2$ can be replaced by $M^{2}_{\pi} ( M^{2}_{K} - M^{2}_{\pi}/2)$.
In this way one recovers the S$\chi$PT-based analysis of 
Ref.~\cite{Moussallam}.
The second term on the right hand side of Eq.~(\ref{delta}) then takes the value
0.21, whereas the evaluation of the sum rule (\ref{sumrule}),
as discussed in Ref.~\cite{Moussallam}, corresponds to the final result
$0.38 < X(2) - X(3) < 0.73$, which is compatible with the conclusion
expressed in Ref.~\cite{Moussallam} in terms of the ratio $X(3)/X(2)$. 
Notice that $\Pi_{Z}(m_s) >0$ for not too large $m_s$, because the two-point
function (\ref{der}) exhibits
a positive logarithmic increase as $m_s \to 0$. Consequently, in 
S$\chi$PT, the bound $X(2)-X(3)>0.21$ must hold up to two-loop corrections.
A too large difference $X(2)-X(3)$ could  hardly be reconciled with the
premises of S$\chi$PT which require both GOR ratios to be reasonably
close to 1. Consequently, a new analysis of the sum rule (\ref{delta}) 
within G$\chi$PT would be highly desirable. 

\section{Summary and concluding remarks}

{\bf 1.} Order parameters of SB$\chi$S which are dominated by the
infrared extremity
of the Euclidean Dirac spectrum ($\Sigma=-\langle \bar{u}u\rangle$,
$F_\pi^2$, \ldots) could be sensitive to the paramagnetic effect of light
quark loops : in the chiral limit, the fermionic determinant reduces the
statistical weight of the lowest Dirac eigenvalues and gradually suppresses
these order parameters as the number of massless flavours $N_f$ increases.
This might result into a rich chiral phase structure as a function of $N_f$
and $N_c$.

{\bf 2.} For $N_f$ approaching the critical point $n_\mathrm{crit}(N_c)$, the
average density of low Dirac eigenvalues should drop and its fluctuations 
should increase. This would naturally lead to a significant reduction of
$\langle \bar{q}q \rangle$, i.e. to a rapid decrease of the
Gell-Mann--Oakes--Renner (GOR) ratio $X(N_f)$ (\ref{GOR}), and to an
enhancement of
the Zweig rule (ZR) violation in the scalar channel as compared to the
large-$N_c$ predictions.

{\bf 3.} In the real world, where
$m_u\sim m_d \ll m_s \ll \Lambda_H \sim 1\ \mathrm{GeV}$, $X(2)$ and $X(3)$
are of a direct phenomenological interest ($1>X(2)>X(3)>0$). If the critical
point (for $N_c$=3) is far away from $N_f=3$, it is natural to expect
$X(3)\sim X(2) \sim 1$ and a less important ZR violation in the
$0^{++}$ channel. If, on the other hand, the real world is close to a phase
transition, $X(N_f)$ should quickly fall towards the critical point,
leading to a large ZR violating difference $X(2)-X(3)$. This would
force a value of $X(3)$ significantly below 1, leaving open the 
question whether $X(2)$ already feels the influence of the critical point
or still remains close to 1.

{\bf 4.} $X(2)$ can be extracted from precise low-energy $\pi\pi$-scattering
experiments, independently of the ZR violation and of the size of
$X(3)$. Furthermore, $X(2)$ is closely related to the quark mass ratio
$r=2m_s/(m_u+m_d)$, and this relation is only marginally affected by $X(3)$.
On the other hand, even if $X(2)\sim 1$, and S$\chi$PT were a reliable
expansion scheme in the two-flavour sector, its accurate predictions for
s-wave scattering lengths \cite{Gasser-Leutwyler,pipistd} would be obstructed by
important ZR
violation already at the one-loop level: sofar, a reliable determination of
the $\mathrm{SU}(2)\times\mathrm{SU}(2)$ low-energy $O(p^4)$ constants
$l_3$ and $l_4$ from independent experimental data requires (among other
things) the knowledge of the $\mathrm{SU}(3)\times\mathrm{SU}(3)$ ZR
violating constants $L_6$ and $L_4$. The determination of the two-flavour GOR
ratio $X(2)$ remains the central goal of ongoing $\pi\pi$ scattering
experiments \cite{daphne,mainz,exper} and of related proposals
\cite{tau3pi,gammagamma}. If $X(2)$ turned out to be close to 1,
these experiments could be interpreted as a first measurement of the
low-energy constants $l_3$ and $l_4$.

{\bf 5.} Whatever the experimental output for $X(2)$ will be, additional
information will be necessary to pin down $X(3)$ and to settle the
theoretical issue of a nearby phase transition. The sum rule analysis of
Ref.~\cite{Moussallam} could be extended, but a more direct access to the ZR
violation in the $0^{++}$ channel and to the three-flavour condensate would
be desirable despite its difficulty.

{\bf 6.} The theoretical question of what happens in the phase in which we
likely do not live [$N_f>n_\mathrm{crit}(N_c)$], is at present hard to
answer unambiguously. The first chiral transition should merely affect
observables that are particularly sensitive to the lowest modes of the Dirac
operator. The $\rho$-meson mass, the string tension and the characteristics
of confinement in general need not be affected at all. Among chiral order
parameters, $\langle \bar{q}q\rangle$ exhibits the strongest infrared
sensitivity and is expected to vanish first. Whether this implies a complete
or partial restoration of chiral symmetry \cite{Appelquist}
is a question that involves the hard problem of non-perturbative regularization 
and renormalization of the chiral symmetry-breaking sector of QCD. Within the
cut-off dependent bare theory, the vanishing of $\langle \bar{q}q\rangle$ implies
$F^2_\pi=0$, and consequently the full restoration of chiral symmetry
\cite{Shifman}. The validity of this argument in the full theory remains to
be clarified. The answer likely resides in a non-perturbative study of the fixed
points of renormalization group flows.

\end{document}